\documentclass[
preprint,
 amsmath,amssymb,
 aps,
showkeys,
showpacs
]{revtex4-2}

\usepackage{comment}
\usepackage{graphicx}
\usepackage{dcolumn}
\usepackage{bm}

\newcommand{\be}{\begin{eqnarray}}
\newcommand{\ee}{\end{eqnarray}}
\newcommand{\bi}{\bibitem}

\begin{document}

\preprint{MUPB/Conference section: }

\title{Antistars in the Galaxy
\\
 }

\author{A.D. Dolgov}
\affiliation{%
 Physics Department, Novosibirsk State University.
 \\
 Bogolyubov Laboratory of Theoretical Physics, 
 JINR, Dubna
}%

\date{\today}

\begin{abstract}
Possible existence  of antimatter in our Galaxy, in particular of antistars is discussed and the mechanism of their
creation is decribed.
\end{abstract}

\maketitle


\section{Baryogenesis and C and CP violation in cosmology}\label{s-bargen}

The excess of matter over antimatter in the universe is explained by the famous 
Sakharov mechanism~\cite{ADS-BG} based on three cornerstones:\\
{$\bullet$ Nonconservation of baryonic number.}\\
{ $\bullet$ C and CP (explicit) violation.}\\
{ $\bullet$ Deviation from thermal equilibrium.}\\
Depending upon the model these  principles allow to calculate the magnitude of the baryonic number density normalised to 
the density of the photons of CMBR:
\be
\beta = N_B/ N_\gamma = const . 
\label{beta}
\ee
In the simplest versions of the scenario no antibaryons are created by the Sakharov mechanism and $\beta$ is 
predicted to be constant over all the universe. There is no room for antimatter in this simple scenario. However, the outcome
depends upon depends upon the mechanism of C and CP breaking in cosmology. They can be separated into three possible
groups~\cite{AD-CP-cosm}:\\
$\bullet$ 1. Explicit, which is achieved by complex coupling constants or masses  in the fundamental Lagrangian of 
particle physics. In this case $\beta= const $ and has fixed sign, so no primordial antimatter is created.\\
$\bullet$ 2. Spontaneous~\cite{lee-spont}, realised by a non-zero expectation value of a (complex) scalar field, which 
can have different signs in different space regions. In this case the universe would contain an equal amount of matter and antimatter 
with matter-antimatter domains separated by cosmologically large distances of the order of Gygaparsecs~\cite{CdRG}. \\
$\bullet$ 3. Dynamical  which operated only in the early universe if a classical complex scalar field somehow 
originated at the early stage of the cosmological evolution, but disappears later leaving no trace. If it happened simultaneously with 
baryogenesis, then the regions with different  values and signs of ${ \beta}$  can be created.
This mechanism left behind neither domain walls nor any effects in particle physics. \\

In versions 2 and 3 stars and antistars can be created but far away from each other, in different galaxies.
None of these, more or less conventional, scenarios allows for abundant antimatter in a galaxy predominantly 
consisting  of matter or vice versa.

\section{Antimatter in the Milky Way \label{s-antimater}}

Possible discovery of several { antistars in the Galaxy} was recently reported~\cite{antistars}. Quoting the authors
"We identify in the catalog 14 antistar candidates not associated with any objects belonging to established gamma-ray 
source classes and with a spectrum compatible with baryon-antibaryon annihilation."

Somewhat earlier a striking idea was put forward that dark matter may consist of compact anti-stars~\cite{anti-DM}.
The authors noted that such anti-DM may be easier to spot than other forms of macroscopic DM. However, detailed 
consideration of the observartional limits is in order.

Bounds on the antistar  density in the Galaxy ware studied in refs.~\cite{anti-1,anti-2,anti-3}. 
As is argued there the fractional density of compact
anti-stars in the universe and even in the Galaxy at the level of about 10\%
does not violate existing observational limits. The relatively weak limit is explained  by the fact that the annihilation
of the interstellar gas with antistar antimatter takes place either on the antistar surface or in the antistellar wind.
Surface annihilation on a compact object is much less efficient than volume annihilation, e.g. inside gas cloud of antimatter.
Such diffuse clouds of antimatter are also predicted by the theory, discussed below, but they could not survive to our
time.

After this Conference another method of antistar identification was suggested~\cite{BBBDP}. It was noticed that prior to 
$p\bar p$-annihilation the Coulomb-bound atomic like state could be formed analogous to positronium. Similar states 
$He \bar p$, $p \bar He$, or $He \bar He$ could also be formed. All such "atoms" were created. at highly excited states
and in the process of de-excitation they would emit narrow X-rays lines with the energies in the range 1-10 keV. Which
might be registered by Roentgen observatories.

\section{Antistar Prediction and  Anti-Creation Mechanism }

Possible existence of antistars in the Galaxy was predicted many years ago in refs.~\cite{AD-JS,DKK}
In these works were a new mechanism of primordial black holes (PBHs) was worked out and 
an appearance of antistars in the Galaxy was a natural by-product of the proposed mechanism.
This mechanism of massive PBH formation predicted in particular the very simple mass spectrum of PBH,  the
so called log-normal mass spectrum:
\be
\frac{dN}{dM} = \mu^2 \exp{[-\gamma \ln^2 (M/M_0)]. }
\label{dn-dM}
\ee
Two out of three constant parameters,  $\mu$ and $\gamma$, are model dependent and cannot be reliably predicted
without information about the properties of high energy particle interactions. However, the value of the central mass $M_0$
should be equal to the mass inside the cosmological horizon at the QCD phase transition which took place in the  
very early universe at the temperatures about 100 MeV. So according to the estimate of  ref.~\cite{M0}
$M_0 \approx 10 M_\odot$. 
The chirp mass distribution of LIGO events very well agrees with the log-normal form of the PBH spectrum~\cite{chirp-mass}. 
This is the only known mass spectrum of PBH verified by observations.
The observed numbers of supermassive black holes ($ M= (10^6 - 10^{10}) M_\odot$, intermediate mass BHs 
$ M= (10^2 - 10^{5}) M_\odot$, and of BHs with masses of tens solar masses are well described by lognormal spectrum.
 The massive PBHs  created by the mechanism of refs.~\cite{AD-JS,DKK} allow to cure multiple inconsistencies 
 related to their origin in the  
the conventional cosmology and astrophysics.  Unusual stellar type compact objects observed in the Galaxy
including abundant antistars in the Galaxy, too quickly moving stars, too old stars, stars with nontypical chemistry~\cite{AD-UFN}.
can also be created by the mechanism \cite{AD-JS,DKK}.
Dark matter made out of PBHs with log-normal spectrum may be a viable option~\cite{carr-extended}.

In short the basic features of the mechanism are the following. It is essentially based on the
SUSY motivated baryogenesis or Affleck-Dine (AD) baryogenesis~\cite{Affl-Dine}. 
SUSY predicts existence of  scalar field $\chi$ with non-zero baryonic naumber, {${ B\neq 0}$.}
Such bosons may condense along the flat directions of the potential of $\chi$. We assume that the potential
contains the quartic and quadratic terms:
\be
U(\chi) = \lambda |\chi|^4 \left( 1- \cos 4\theta \right) + m^2 |\chi|^2} {\left[{ 1-\cos (2\theta+2\alpha)} \right],
\ee
where ${ \chi = |\chi| \exp (i\theta)}$ and ${ m=|m|e^\alpha}$.
{If ${\alpha \neq 0}$, C and CP are  broken.} 
In GUT SUSY baryonic number is naturally non-conserved. Thus is expressed through the non-invariance of ${U(\chi)}$
with respect to the phase rotation of $\chi$. 

{Initially, just after inflation, the classical field ${\chi(t)}$ was away from the origin and, when 
inflation was over, it started to evolve down to the equilibrium point, ${\chi =0}$,
according to the equation analogous to that of the Newtonian mechanics with liquid (Hubble) friction:}
\be
\ddot \chi +3H\dot \chi +U' (\chi) = 0.
\label{ddot-chi}
\ee
Baryonic number of $ \chi (t)$, equal to $B_\chi =\dot\theta |\chi|^2$,
is analogous to mechanical angular momentum.  At some later epoch the decay of
${{\chi}}$ transferred its
baryonic number to that of quarks in B-conserving process.
In contrast to other baryogenesis scenarios, the
{AD baryogenesis could lead to baryon asymmetry of order of unity, much larger
than the observed  value  $ \beta \sim {10^{-9}}$.} If necessary, with mild modification, the baryon asymmetry can be reduced to the 
a much smaller value.

If ${ m\neq 0}$, the angular momentum, $B_\chi$, could be generated by different 
directions of the  quartic and quadratic valleys at low ${\chi}$.
{If CP-odd phase ${\alpha}$ is small but non-vanishing, both baryonic and 
antibaryonic domains might be  formed}
{with possible dominance of one of them.}\\
{Matter and antimatter domains may exist but globally ${ B\neq 0}$.}

The essential feature of the mechanism~\cite{AD-JS,DKK}  is the introduction of the 
general renormalizable coupling of $\chi$ field to 
the inflaton $ \Phi$, the first term in the equation below:
\be 
U = {g|\chi|^2 (\Phi -\Phi_1)^2}  +
\lambda |\chi|^4 \,\ln \left( \frac{|\chi|^2 }{\sigma^2} \right)
+\lambda_1 (\chi^4 + h.c. ) + 
(m^2 \chi^2 + h.c.),
\ee
where $\Phi_1$ is the value which $\Phi$ passed during inflation, some time before its end. The logarithmic term in this 
expression is a result of one-loop contribution to the potential, i.e. the Coleman-Weinberg potential~\cite{C-W-pot}.

When $\Phi$ is close to  $\Phi_1$, the window to the flat direction is open and 
{the field ${\chi}$ can diffuse to a large value.} 
If the window to flat direction 
is open only during a sufficiently short period, cosmologically small but possibly astronomically large 
bubbles with high ${ \beta}$ could be created, occupying {a small
fraction of the universe,} while the rest of the universe has normal
${{ \beta \approx 6\cdot 10^{-10}}}$, created 
by small ${\chi}$. 

{This mechanism of massive PBH formation is quite different from all others.}
{The fundament of PBH creation is build at inflation by making large isocurvature
fluctuations at relatively small scales, with practically vanishing density perturbations.} 
{Initial isocurvature perturbations are contained in a large excess of quarks over antiquarks (or vice versa).
Since quarks at this stage are massless the density perturbations are practically absent.
Density perturbations are generated rather late after the QCD phase transition.}
{The emerging universe looks like a piece of Swiss cheese, where holes are high baryonic 
density objects occupying a minor fraction of the universe volume.} Since $\chi$ could rotate in the complex $\chi$-plane
clockwise or anticlockwise around the origin, both bubbles with high baryonic or antibaryonic numbers could be created.
Mostly they would turn into PBHs but those which are not massive enough remains compact stars or antistars,

A competing suggestion on galactic antimatter was proposed recently in ref.~\cite{zb}.
According to the statement of the author
the oscillation of the neutron $n$ into mirror neutron $n’$, its partner from dark mirror sector, can gradually transform into
an ordinary neutron star into a mixed star consisting in part of mirror dark matter. The implications of the reverse process taking place in 
the mirror neutron stars depend on the sign of baryon asymmetry in mirror sector. Namely, if it is negative, as predicted by certain 
baryogenesis scenarios, then 
{$\bar{n}’-\bar{n}$ { transitions create a core of our antimatter gravitationally trapped in 
the mirror star interior.}}
The annihilation of accreted gas on such antimatter cores could explain the origin $\gamma$-source candidates, with unusual spectrum 
compatible to baryon-antibaryon annihilation~\cite{antistars}, 
after the mergers of mirror neutron stars can produce the flux of cosmic antihelium and also heavier antinuclei which are hunted in the 
AMS-02 experiment.

\begin{acknowledgments}
This work was supported by the RSF Grant 20-42-09010.
\end{acknowledgments}

\nocite{*}


\end{document}